\documentclass[english,superscriptaddress]{revtex4}
\usepackage[T1]{fontenc}
\usepackage[latin9]{inputenc}
\usepackage{amsmath}
\usepackage{babel}
\usepackage{color}
\usepackage{subfigure}
\usepackage{epsfig}
\usepackage[letterpaper, margin=1in]{geometry}

\usepackage{lmodern,enumerate,natbib}
\usepackage{amssymb,amsthm,amsfonts,mathrsfs}

\newcommand{\comm}[2]{[#1, #2]}

\newcommand{\lgen}{\mathcal{L}}
\newcommand{\usup}{\mathcal{U}}

\begin{document}
\author{Robert Alicki}
\thanks{These authors contributed equally to this work.}
\affiliation{Institute of Theoretical Physics and Astrophysics, University of Gdansk, Wita Stwosza 57, 80-952 Gdansk, Poland}

\author{David Gelbwaser-Klimovsky}
\thanks{These authors contributed equally to this work.}
\affiliation{Department of Chemistry and Chemical Biology, Harvard University,
Cambridge, MA 02138}

\title{Non-equilibrium quantum heat machines }

\begin{abstract}
Standard heat machines (engine, heat pump,  refrigerator) are composed of  a system (``working fluid") coupled to at least two equilibrium baths at different temperatures and periodically driven by an external device (piston or rotor) called sometimes work reservoir. The aim of this paper is to go beyond this scheme by considering environments which are stationary but cannot be decomposed into few baths at thermal equilibrium. Such situations are important, for example in  solar cells, chemical machines in biology, various realizations of laser cooling or  nanoscopic machines 
driven by laser radiation. We classify non-equilibrium  baths depending on their thermodynamic behavior and show that the efficiency of heat machines operating under their influences is limited by a generalized Carnot bound.
\end{abstract}
\maketitle

% es solo para fast driving?

\section{Introduction}

Quantum systems are rarely completely isolated from their environment, whose influence, positive or negative, should be considered.   The theory of open quantum system was developed \cite{Davies:1974,Lindblad76,AlickiLendi:2006,HuelgaRivas:2012,spohn2007irreversible} to achieve this goal and in particular  open the way to the study of quantum heat machines, such as engines and refrigerators \cite{gelbwaser2015thermodynamics,kosloff2013quantum,segal2006molecular,esposito2010quantum,Levy:2012,correa2013performance}.  Those models generally assume the interaction between a system and one or two environments,  in thermal equilibrium, thereby termed heat baths. The efficiency of these machines is limited by the Carnot bound, requiring at  least two baths at different temperatures in order to extract work.
\par
 Nevertheless, there are many examples in nature where the environment is not in thermal equilibrium, such as sunlight, continuous laser radiation, biological cells, etc. Our goal is to establish maximum efficiency bounds, as well as to determine the output (work power or cooling power) of quantum heat machines operating with  non-equilibrium baths.  Because the bath is not at thermal equilibrium, the second law does not demand the presence of a second bath for work extraction.  Our aim is to go beyond a simple situation when an environment is a collection of independent heat baths with different temperatures which, in principle can be also treated as a single non-equilibrium bath.   
\par
We study a micro or mesoscopic externally driven quantum system, the working fluid, coupled to a large environment. On the relevant time-scale, which is longer than the scales of driving and of microscopic irreversible processes, the basic parameters of the reservoir are constant, hence  the reference state of the environment is a stationary state. Such non-equlibrium stationary systems are well known in macroscopic thermodynamics  and usually described in terms of local equilibrium with space-dependent temperature, density, pressure, fluid velocity etc \cite{kondepudi2014modern}. 
\par
However, there are important situations where  non-equilibrium character of the environment state is not related to spatial non-homogeneity but rather to some internal properties of its state. For example, sunlight at the Earth surface is a rather homogeneous environment, the shape of its spectrum roughly corresponds to the Planck distribution at the Sun surface temperature $T_s$, but the photon density is much lower than the equilibrium one,   and the absorption in the atmosphere creates ``holes'' in the spectrum. 
\par
Another example is a laser radiation \cite{scully1997quantum} in a continuous wave operation mode, which for the idealized single mode situation can be treated  as a single quantum oscillator or even a classical monochromatic wave, while for multimode case with strong phase diffusion it acts on optically active centers like a non-equilibrium bath. Biological machines also provide  examples of systems coupled to non-equilibrium baths either of the chemical nature or consisting of photons or different types of excitons.

In the next Section we briefly review the theory of quantum heat machines operating with equilibrium baths and consider the simplest case, where the working fluid is a two-level system (TLS). In Section III the non-equilibrium  TLS heat machine is analyzed. The notion of ``local temperatures'' is introduced. Different examples of non-equilibrium baths are given and they are classified according to their effects on the heat machine operation. Finally, in Section IV a general theory for non-equilibrium quantum heat machines composed of a general quantum working fluid is given and showed that their maximum efficiency is limited by a Carnot-like bound.

\section{Standard equilibrium  heat machines}

As a first step and reference point we review the main results of
quantum heat machines operating in contact with two equilibrium baths.
The machine is composed of a periodic modulated working fluid, the system, which is
permanently coupled to the hot (cold) bath at temperature $T_{H(C)}$,
and follows a continuous cyclic evolution \cite{Gelbwaser:2013,SzczygielskiGelbwaserAlicki:2013}. The total Hamiltonian is 
\begin{equation}
H_{tot}(t) = H_{S}(t)+\sum_{i= H,C} \bigl(S_{i}\otimes B_{i}+H_{B_{i}}\bigr) ,
\label{eq:hamtot}
\end{equation}
where $H_{B_{i}}$ is the $i$-bath free Hamiltonian, $B_{i}$ its interaction
operator and $S_{i}$ is a system operator. The system Hamiltonian fulfills
 periodicity condition $H_{S}(t)=H_{S}(t+\frac{2\pi}{\Omega})$,
$\Omega$ being the modulation angular frequency.

To illustrate our approach, we choose the simplest realization as an example, nevertheless the same analysis may be applied to more complex models. We
assume the working fluid is a TLS whose frequency is modulated,
$H_{S}(t)=(\omega_{o}+\omega(t))\frac{\sigma_{Z}}{2}$. The modulation
may be decomposed into a Fourier series $e^{-i\int_{0}^{t}\omega(t')dt}=\sum\xi_{q}e^{-iq\Omega t}$ 
leading to the Floquet expansion of the Lindblad operator \cite{SzczygielskiGelbwaserAlicki:2013}. 

For a detailed derivation we refer the reader to  \cite{Szczygielski:2014}. In the interaction
picture the evolution equation of the working fluid density matrix, $\rho$, is given by the quantum Markovian master equation of Lindblad-Gorini-Kossakowaski-Sudarshan type \cite{GKS76,Lindblad76}, which for diagonal matrix elements (${\rho}_{gg(ee)}$, ground (excited) state population) yields the rate equation

\[
\dot{\rho}_{ee}=-\sum_{q\in Z}\sum_{i= H,C} \left(P_{q}G^{i}(\omega_{q})\rho_{ee}+P_{q}G^{i}(-\omega_{q})\rho_{gg} \right),
\]
where $\omega_q=\omega_0+q\Omega$, $P_{q}=||\xi_{q}||^{2}$ are the harmonic strength satisfying normalization condition $\sum_{q\in Z}P_{q} = 1$. The bath
coupling spectrum is defined as $G^{i}(\omega)=\int_{-\infty}^{\infty}e^{it\omega}\langle B^{i}(t)B^{i}(0)\rangle dt$
and measures the interaction strength between the TLS and the bath mode $\omega$.
If the bath is at thermal equilibrium, it fulfills  the Kubo-Martin-Schwinger (KMS) condition (we put $h=1$ and $k_B=1$) \cite{kubo1957statistical,martin1959theory}
:
\begin{equation}
\frac{G^{i}(-\omega)}{G^{i}(\omega)}=e^{-\omega/T_{i}},\label{eq:kms}
\end{equation}
where $T_{i}$ is the bath temperature. If the TLS interacts only with the i-bath and is not being modulated it will equilibrate to this temperature, i.e, $\rho_{ee}/\rho_{gg}=e^{-\omega_{0}/T_i}$.

We are interested in the thermodynamic behavior at steady state (or
limit cycle) where any transient effect vanishes. The steady state
heat currents from the cold and hot bath, respectively, and the power supplied by the source of modulation  (work reservoir) are given by the expressions 

\begin{gather}
\bar{J}_{C(H)}=\sum_{q \in \mathbb{Z}}\frac{\omega_{q}P_{q}}{w+1}\left(G^{C(H)}(-\omega_{q})-G^{C(H)}(\omega_{q})w\right), \notag\\
\bar{P}=-\bar{J}_{H}-\bar{J}_{C},
\end{gather}
\\
where $w$ is the steady state population rate
\[
w=\left(\frac{\rho_{ee}}{\rho_{gg}}\right)^{SS}=\frac{\sum_{q \in \mathbb{Z}}\sum_{i = H,C}P_{q}G^{i}(-\omega_{q})}{\sum_{q \in \mathbb{Z}}\sum_{i = H,C}P_{q}G^{i}(\omega_{q})}
\]
and the sign convention is that extracted power is negative.

As expected from the Second Law, it has been shown \cite{Gelbwaser:2013,Gelbwaser:2013a} that work extraction requires at
least two equilibrium baths at different temperatures. Nevertheless, this does not apply to non-equilibrium baths. In the following sections
we  address this scenario and find the conditions needed for extracting work from a single non-equilibrum bath.

\section{How to introduce the formalism of non-equilibrium baths}
We consider the case of a TLS coupled to a  \textit{single}  stationary non-equilibrium bath \cite{kondepudi2014modern} through the interaction Hamiltonian $H_{int}=S\otimes B$. Stationarity requires the bath state to be  diagonal in the bath  free Hamiltonian basis. Thermal states are just  particular cases of this kind of states.  Stationarity is essential to ensure that  the bath two-times auto-correlations depend only on time difference.

In order to describe non-thermal but stationary baths,  we generalize the  KMS condition \eqref{eq:kms}, by introducing 
``local temperature''%
\begin{gather}
e^{-\omega/T_B(\omega)}  \equiv \frac{G(-\omega)}{G(\omega)},
\label{eq:nokms}
\end{gather}
where $B$ is the bath interaction operator. Local temperatures depend on $B$, the frequency and  the state of the bath. Only for a thermal  equilibrium bath $T_B(\omega) = T$ independently of $B$ and $\omega$.

In the next section we show some examples of non-equilibrium baths and calculate their effective local temperature.

\subsection{Harmonic oscillator baths}

Baths like  electromagnetic radiation or  vibration modes of a material are  just a collection of independent quantum oscillators with quasi-continuous spectrum of frequencies \cite{breuer2002theory}. Baths could also be composed of fermions, e.g., spin-baths, but even then by suitable transformations (e.g. Holstein-Primakoff  Hamiltonian) can often be approximated by bosonic baths. The bosonic bath free Hamiltonian is given by 
\begin{equation}
H_B = \sum_{k} \omega_k\, b_k^+ b_k , \quad  [b_k , b_l^+] = \delta_{kl}.
\label{harmonic_bath}
\end{equation} 
In most applications the bath operator that couples to the system  is linear in creation and annihilation operators
\begin{equation}
B= \sum_{k} \bigl\{ g_k b_k + \bar{g}_k b_k^+\bigr\}.
\label{harmonic_bath1}
\end{equation} 
The state of the bath is assumed to be stationary and hence diagonal in the particle number basis.
Then, the coupling spectrum yields %
\begin{equation}
G(\omega) = \left\{ \begin{array}{ll}
	\sum_{k} |g_k |^2 ( n_k + 1)\delta(\omega_k - \omega), & \omega > 0 \\
	\sum_{k} |g_k |^2  n_k \delta(\omega_k - \omega), & \omega < 0 
	\end{array}\right\},
\label{harmonic_spec}
\end{equation} 
where $n_k = \mathrm{Tr}(\rho_B b_k^+ b_k)$ is the k-mode population, the upper (lower) line in \eqref{harmonic_spec} is the emission (absorption) rate. The local temperature is given by
\begin{equation}
T_B(\omega) = \frac{\omega}{\ln\bigl[ \frac{G(\omega)}{G(-\omega)} \bigr]}
\label{linear_temp}
\end{equation} 
or
\begin{gather}
e^{-\omega/T_B(\omega)} = \frac{n(\omega)}{n(\omega) +1},
\label{eq:nokms1}
\end{gather}
where $n(\omega)= \frac{\sum_{k} |g_k |^2  n_k \delta(\omega_k - \omega)}{\sum_{k} |g_k |^2 \delta(\omega_k - \omega)}$ denotes the average population number for the frequency $\omega$.

As we show below, the frequency dependence of the local temperature determines whether work can be extracted from the single non-equilibrium bath.

\subsubsection{Sunlight}
\begin{figure}
	\centering
		\includegraphics[width=0.85\textwidth]{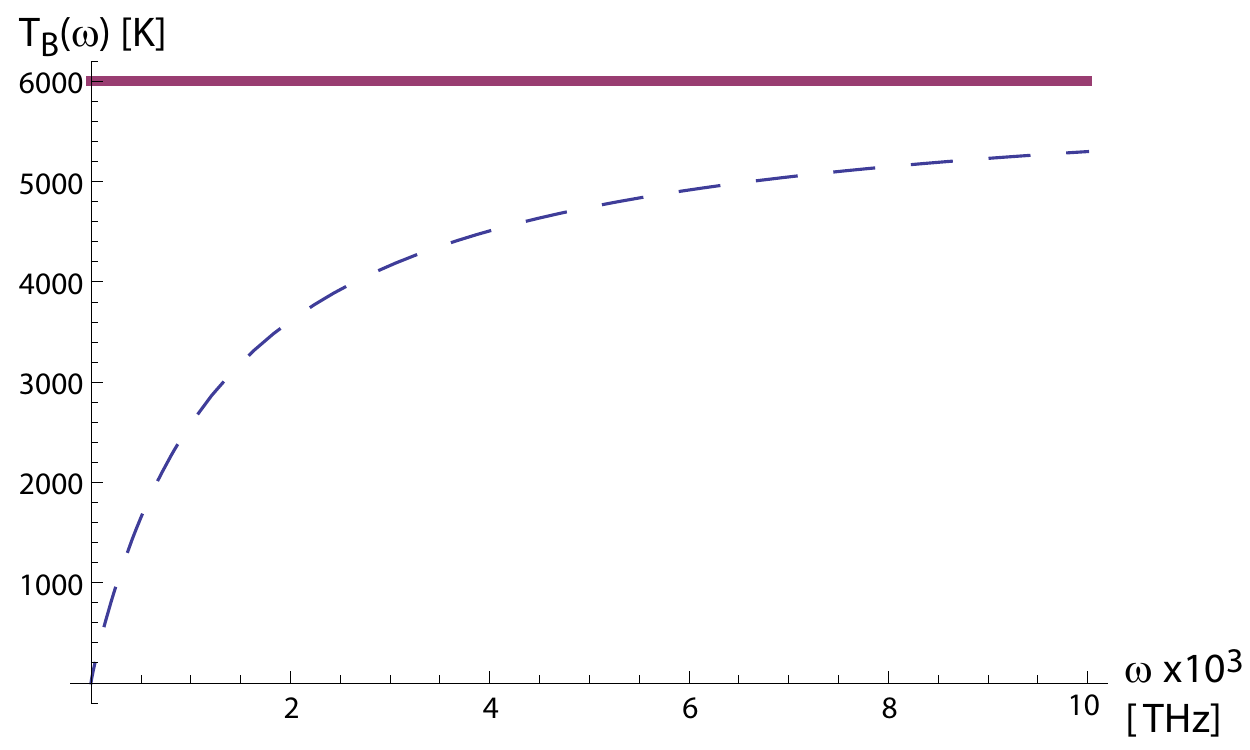}
		\caption{(Color online) Local temperature of a TLS coupled to  sunlight. The continuous line corresponds to the case of thermal radiation at the temperature of Sun surface $T_s \simeq 6000 K$, while the dashed line to sunlight out of equilibrium due to a geometrical factor,  $\lambda= 2.5\times 10^{-5}$  (see main text). }
	\label{fig:C}
\end{figure}

The Sun is a thermal source, emitting thermal radiation at $T_s= 6000k$. Due to geometrical considerations \cite{wurfel2009physics,alicki2015solar}, just a small fraction of the emitted photons reach the Earth, reducing the effective mode population  and thereby  $n(\omega) = \lambda \bigl[e^{\omega/T_s} -1\bigr]^{-1}$, where $\lambda= 2.5\times 10^{-5}$ is a geometric factor equal to the angle subtended by the Sun seen from the Earth.  Effectively, sunlight on Earth is out of equilibrium, and systems with different frequencies, will ``equilibrate'' to different temperatures.  Fig \ref{fig:C}, shows the equilibration temperature, $\rho_{ee}/\rho_{gg}=e^{- \omega/T_B(\omega)}$ as a function of the TLS frequency. Moreover, the atmosphere acts like a filter and produces a more complicated shape of $n(\omega)$ with many ``holes". 

\subsubsection{Multimode laser radiation}
A multimode laser radiation  in a continuous wave operation mode may be modeled as a   bath at the phase average,  $\bar{\rho}_B$, of the multimode coherent state $\rho_B$, 
\begin{equation}
 \rho_B=U_{z} |\mathrm{vac}\rangle\langle \mathrm{vac}| U_{z}^{\dagger}, \quad |\mathrm{vac}\rangle\langle \mathrm{vac}|= \bigotimes^k |0_k\rangle\langle 0_k|,
\label{coherent}
\end{equation} 
where $U_{z}^{\dagger}$ is the displacement operator
\begin{equation}
U_{z}^{\dagger} b_k  U_{z} = b_k - z(\omega_k).
\label{coherent1}
\end{equation} 
This non-equilbrium bath is obtained by displacing a thermal equilibrium bath at zero temperature and performing a phase averaging.
The phase average of the bath state, or diagonality in the photon occupation number basis, is required in order for the bath to be stationary. 
The local temperature  for this bath is  equal to
\begin{equation}
T_B(\omega) = \frac{\omega }{\ln\bigl(1 + |z(\omega)|^{-2}\bigr)} .
\label{coherent2}
\end{equation} 
For large $|z(\omega)|$ yields
\begin{equation}
T_B(\omega) \simeq \omega |z(\omega)|^{2} .
\label{coherent3}
\end{equation}

Therefore, for a constant shift, $z$, for all the modes, the local temperature is a linearly increasing function of the frequency. 

\subsubsection{Squeezed thermal bath}
The state of a stationary squeezed thermal bath  is the phase average \cite{abah2014efficiency,rossnagel2014nanoscale}, $\bar{\rho}_B$, of the following density matrix

\begin{equation}
 \rho_B = Z^{-1}S_{r}\, e^{- H_B/T_{eq}}\, S_{r}^{\dagger},
\label{squeezed}
\end{equation} 
where $S_r$ is the squeezing unitary operator defined by
\begin{equation}
S_{r}^{\dagger} b_k  S_{r} = \cosh(r(\omega)) b_k + \sinh(r(\omega)) b_k^+ .
\label{squeezed1}
\end{equation} 
Its  local temperature is given by the expression
\begin{equation}
T_B(\omega) = \frac{\omega }{\ln\bigl\{1 + \bigl[n_{T_{eq}}(\omega) + (2 n_{T_{eq}}(\omega) +1)\sinh^2(r(\omega))\bigr]^{-1}\bigr\}} ,
\label{squeezed2}
\end{equation} 

where  $n_{T_{eq}}(\omega) = [e^{\omega/T_{eq}} -1]^{-1} $, is the mode population without squeezing. For large $r$  \eqref{squeezed2} reduces to
\begin{equation}
T_B(\omega) \simeq \frac{\omega }{4} (2n_{T_{eq}}(\omega) +1) e^{2r} .
\label{squeezed3}
\end{equation}

\subsection{Classification of non equilibrium baths}

The  functional dependence of $T_B(\omega)$ can be used to classify
 non-equilibrium baths. It depends on the bath state, coupling to the system and frequency. In this sections we consider
a heat engine composed of a TLS coupled to a single non-equilibrium
bath and find the required conditions   for work extraction. The dynamics of the total system is governed by the Hamiltonian 
\begin{equation}
H_{tot}(t) = (\omega_{o}+\omega(t))\frac{\sigma_{Z}}{2}+ S\otimes B + H_{B}.
\label{eq:hamtot2}
\end{equation}
The steady heat current is a sum of contributions corresponding to all harmonics of $\Omega$ (see section IV for a general case)
\begin{equation}
\bar{J}=\sum_{q \in \mathbb{Z}} \frac{\omega_{q}P_{q}}{w+1}\left(G(-\omega_{q})-G(\omega_{q})w\right) = -\bar{P},
\label{eq:curstd}
\end{equation}
with  the  TLS steady state population ratio
\begin{equation}
w=\left(\frac{\rho_{ee}}{\rho_{gg}}\right)^{SS}=\frac{\sum_{q \in \mathbb{Z}}P_{q}G(-\omega_{q})}{\sum_{q \in \mathbb{Z}}P_{q}G(\omega_{q})}.
\label{eq:w}
\end{equation}
\subsubsection{Non-equilibrium but passive }
Using Eqs \ref{eq:nokms},\eqref{eq:curstd}, \eqref{eq:w} we can write the expression for power supplied to the system as 
\begin{equation}
\bar{P}=  z^{-1}\sum_{\{q_{1}>q_{2}\in \mathbb{Z}\}}{(q_{1}-q_{2})\Omega P_{q_{1}}P_{q_{2}}G(\omega_{q_1})G(\omega_{q_2})}\left(e^{-\omega_{q_2}/T_B(\omega_{q_2})}-e^{-\omega_{q_1}/T_B(\omega_{q_1})}\right),
\label{eq:ppasive}
\end{equation}
where
\begin{equation}
z =\sum_{q \in \mathbb{Z}}P_{q}G(\omega_{q})\bigl[ 1+ e^{-\omega_{q}/T_B(\omega_{q})}\bigr] .
\label{eq:z}
\end{equation}
A sufficient condition 
\begin{equation}
\omega_{q_2}/T_B(\omega_{q_2})>\omega_{q_1}/T_B(\omega_{q_1}), \quad \mathrm{for} \quad \{q_{1}>q_{2}\},
\label{eq:workcond}
\end{equation}
assures that $\bar{P} < 0$ and hence the engine extracts work from the bath.
\par
We define the \textit{passivity function} as $f(\omega) \equiv \frac{d}{d\omega} \left(\omega/T_B(\omega) \right)$. If for all frequencies  $f(\omega)>0$, no work can be extracted. We term such couplings to bath \textit{passive} (in analogy to passive states  \cite{pusz1978passive,lenard1978thermodynamical,Gelbwaser:2013a,gelbwaser2014heat} which do not allow
for work extraction).  Previous examples, based on linear coupling to bosonic bath, are passive if we consider only  frequency independent deformations of a thermal bath  (constant filtering, displacement or squeezing). From Eq \ref{eq:ppasive} we deduce that work
extracted from a single non-equilibrium bath depends on the coupling spectrum shape and requires the passivity function $f(\omega)$
to be negative  in some range of  frequencies.

\subsubsection{Two equilibrium baths as a single non-equilibrium bath}

As a first example of  a non passive bath, we consider the standard quantum heat engine where the working fluid interacts
with two  baths at equilibrium. We model it as a quantum heat machine with a single non-equilibrium bath. For this we need
to consider the spectrum as a sum of  two baths spectra. Therefore,
the local  temperature of the composed bath satisfies the following relation
\[
e^{-\omega/T_B(\omega)}=\frac{e^{-\omega/T_h}G^{h}(\omega)}{G^{h}(\omega)+G^{c}(\omega)}+\frac{e^{-\omega/T_c}G^{c}(\omega)}{G^{h}(\omega)+G^{c}(\omega)}=e^{-\omega/T_h}(1-m(\omega))+e^{-\omega/T_c}m(\omega),\quad 0\leq m(\omega)\leq 1 ,
\]
where we use the fact that both baths are in equilibrium and the standard
KMS condition holds. The effective Boltzmann factor is a weighted average of
both two Boltzmann ones, where the weights depend on how strong is the
working fluid coupled to each bath at the given frequency. Therefore, $T_h\geq T_B(\omega)\geq T_c$. For the sake of simplicity, we assume that the bath coupling spectrum overlaps with only two
harmonic frequencies ($\omega_{q_{1}}>\omega_{q_{2}}$).
As shown in \cite{Gelbwaser:2013} this condition is required in order to achieve high efficiency.
Work extraction requires the hot bath being coupled more strongly
to the high  frequency  mode ($G^{h}(\omega_{q_1})\gg G^{c}(\omega_{q_1})$)
and the opposite for the low  frequency mode ($G^{h}(\omega_{q_2})\ll G^{c}(\omega_{q_2})$) \cite{Gelbwaser:2013}.
The efficiency of the engine is given by the Carnot-type formula
\[
\eta=1-\frac{T_B(\omega_{q_2})}{T_B(\omega_{q_1})}\leq1-\frac{T_c}{T_h}.
\]
The extreme case, where the engine reaches the maximum efficiency,
the Carnot bound, is when the bath are spectrally  separated,  $T_B(\omega_{q_1})\simeq T_{h}$
and $T_B(\omega_{q_2})\simeq T_{c}$.

\subsubsection{Non-equilibrium and non-passive bosonic bath}

As shown above, frequency independent deformation of a thermal bosonic bath with linear coupling, creates passive baths. Therefore, work extraction requires the use of \textit{selective} filters. In order to not contradict the second law of thermodynamics, which
forbids work extraction from a single thermal bath, this selective
filter should involve the presence of other bath, a non equilibrium
process or a hidden work injection.

Assume a  single bosonic bath at the equilibrium temperature $T_{eq}$ and linearly coupled to the system. If a  selective filter, $\lambda(\omega)$, is applied, the local temperature satisfies 
\begin{equation}
e^{-\omega/T_B(\omega)}=\frac{\lambda(\omega)n(\omega)}{\lambda(\omega) n(\omega)+1}.
\end{equation}
What are the conditions required for this filter to allow work
extraction?  
\par
Consider again the bath coupling spectrum which overlaps with only two
harmonic frequencies $\omega_{q_1},  \omega_{q_2}$ with $q_1 > q_2$.  As shown in \cite{Gelbwaser:2013} this condition is required in order to achieve high efficiency and due to \eqref{eq:ppasive} the power is given by
\begin{equation}
\bar{P}= z^{-1}(q_{1}-q_{2})\Omega P_{q_{1}}P_{q_{2}}G(\omega_{q_1})G(\omega_{q_2})\left(e^{-\omega_{q_2}/T_B(\omega_{q_2})}-e^{-\omega_{q_1}/T_B(\omega_{q_1})}\right).
\label{eq:pnopasive}
\end{equation}
Assume that we  reduce the population of the mode $\omega_{q_2}$
($\lambda(\omega_{q_2})$<1), and we do not filter the other
mode ($\lambda(\omega_{q_1})=1)$. In order to allow work extraction, the filter should satisfy the following condition
\[
\lambda(\omega_{q_2})<e^{\frac{(q_{2}-q_{1})}{2T_{eq}}\Omega}\frac{\sinh(\frac{\omega_{q_2}}{2T_{eq}})}{\sinh(\frac{\omega_{q_1}}{2T_{eq}})}<1 .
\]
As paradoxical it may sound, by reducing  a specific
mode population, we can extract work from a single thermal bath! The filtering lowers the effective temperature $T_B(\omega_{q_2})$, reducing the excitations that are emitted to this mode, an  ``saving'' energy which is ultimately transformed into work.
The efficiency of this machine is bounded by a generalized Carnot
limit
\[
\eta=\frac{-\bar{P}}{\bar{J}_{H}}\leq1-\frac{T_B(\omega_{q_2})}{T_{eq}},
\]
where $T_{eq}$ and $T_B(\omega_{q_2})$ play the
role of the effective hot and cold bath temperature respectively.

\subsubsection{Deviation from equilibrium of engineered bath}
Using the selective filtering introduced in the last section, we  can engineer a non-equilibrium bath from an equilibrium one, and characterize its deviation from equilibrium by the parameter
\[
D=1-\lambda(\omega_{q_2}).
\]
For  the  equilibrium bath  $D=0$, and when we start reducing 
population  of the modes with frequencies around $\omega_{q_2}$, the bath will go away from equilibrium
producing some non-equilibrium effects, like frequency dependent equilibration
temperature. Nevertheless work extraction will be possible only when the bath is far enough from equilibrium, i.e.
\begin{equation}
D>n(\omega_{q_1})(e^{\omega_{q_1}/T_{eq}}-e^{\omega_{q_2}/T_{eq}}).
\label{eq:distwork}
\end{equation}
When the deviation from equilibrium increases, the local temperature of
the lower frequency mode reduces and the efficiency of the quantum heat engine rises.
For $D<0$ the bath is also out of equilibrium. But, in this case
instead of reducing the mode population, it is being increased by some
external mechanism (for example, selective concentration of light). The equilibration temperature of the system will
depend on the frequency, but for $D<0$, work cannot be extracted.
Again, paradoxically, the increase of energy in  
``incorrect modes'' reduces the possibility of work extraction.
The bath is taken away from thermal equilibrium, but in the ``opposite direction''
to that leading to work extraction.

\section{General theory for non-equilibrium quantum heat machines}

The model based on the TLS and studied above is an example of a large class of open quantum systems  with  physical Hamiltonian  $H_S(t) = H_S(t +\tau)$ under the assumption that the perturbation frequency $\Omega = 2\pi/\tau$ is comparable to or higher than the relevant Bohr frequencies. Such fast driving is typically provided by a strong coherent laser field and appears in the thermodynamical approach to the theory of lasers  \cite{scully1997quantum,scovil1959three,geva1996quantum,boukobza2007three}, various types of laser cooling \cite{phillips1998nobel,gelbwaser2015laser}, optomechanical devices \cite{aspelmeyer2014cavity}, etc.
Potential applications include new  light harvesting systems, both of  biological nature or man-made devices. For this class of systems a consistent theory can be developed which includes the general case of stationary nonequilibrium environment characterized by local temperatures. The laws of thermodynamics can be derived and Carnot-type bounds are obtained. 
\subsection{Master equations}
We begin with a presentation of the canonical construction of the Markovian  generator for an open system weakly coupled to a stationary, but generally nonequilibrium, environment.
The system is  assumed to be ``small" and described by the periodic in time physical Hamiltonian  $H_S(t) = H_S(t +\tau)$ under the assumption that the perturbation frequency $\Omega = 2\pi/\tau$ is comparable to or higher than the relevant temporal Bohr frequencies. We assume that the Hamiltonian of the system already contains all Lamb-like shifts induced by interaction with environment \cite{AlickiLendi:2006,HuelgaRivas:2012,Szczygielski:2014}.\\
The system bath-interaction is parametrized as
\begin{equation}
	H_{\mathrm{int}} = \sum_{\alpha } S_{\alpha} \otimes B_{\alpha},
\label{Hint}
\end{equation}
where $S_{\alpha}$ and $B_{\alpha}$ are hermitian operators of  system and bath, respectively. The environment (bath) is assumed to be a large quantum system with practically continuous Hamiltonian spectrum and  a proper behavior of multi-time correlation functions of relevant observables. The initial state of the bath  is stationary, therefore is invariant with respect to the free dynamics of the bath and satisfies $\langle B_{\alpha}\rangle_B =0$, where  $\langle\cdots\rangle_B$ denotes the average over the bath state. We assume also, for simplicity, that  the cross-correlations between  $B_{\alpha}$ and $B_{\beta}$ vanish for $\alpha\neq\beta$.
\par
Applying the Floquet theory one obtains the following decomposition  of the associated unitary propagator
\begin{equation}\label{eq_PropagatorResolution}
	U(t) = \mathbb{T} \exp\Bigl\{-i\int_0^t H_S(s) ds\Bigr\} =  P(t) e^{-i\bar{H}t},
\end{equation}
where $P(t) = P(t+\tau)$ is a family of periodic unitaries and $\bar{H}$ is the \emph{averaged Hamiltonian}  satisfying
\begin{equation}\label{eq_FloquetOperatorGeneral}
	U(\tau) = e^{-i\bar{H}\tau} .
\end{equation}
The \emph{Floquet operator} $U(\tau)$, and the averaged Hamiltonian $\bar{H}$  posses  common eigenvectors $\{\phi_{k}\}$, i.e.,
\begin{equation}\label{eq_Floqueteigen}
\bar{H} \phi_{k} = \bar{\epsilon}_k \phi_{k} ,\quad U(\tau)\phi_{k} = e^{-i\bar{\epsilon}_{k}\tau}\phi_{k},
\end{equation}
where $\{\epsilon_{k}\}$ are called \emph{quasi-energies} of the system. These properties imply a particular form of the Fourier decomposition 
\begin{equation}\label{eq_SoperatorExpansion}
	S_{\alpha}(t) \equiv U(t)^{\dagger} S_{\alpha} U(t) = \sum_{\{\bar{\omega}_q\}}S_{\alpha}(\bar{\omega}_q) e^{-it\bar{\omega}_q}, \\
\end{equation}
where
\begin{equation}\label{quasiBohr}
\{\bar{\omega}_q\} = \{ \bar{\omega} + q\Omega\}\, ;\, \{\bar{\omega}\} = \{\bar{\epsilon}_k - \bar{\epsilon}_l\} , q\in\mathbb{Z}\},
\end{equation}
i.e., it is a set of all sums of the \emph{relevant Bohr quasi-frequencies} and all multiplicities of the modulation frequency. By $\{\bar{\omega}_q\}_+$ we denote the subset of $\{\bar{\omega}_q\}$ with non-negative relevant Bohr quasi-frequencies $\bar{\omega}$.
\par
The operators $S_{\alpha}(\bar{\omega}_q)$  are subject to relations 
\begin{align}\label{SrelationsP}
S_{\alpha}(\bar{\omega}_q)^{\dagger} &= S_{\alpha}(-\bar{\omega}_q), \nonumber\\ [\bar{H}, S_{\alpha}(\bar{\omega}_q)] &= -\bar{\omega} S_{\alpha}(\bar{\omega}_q).
\end{align}
Physically, the  harmonics $ q\Omega$ correspond to the energy quanta   which are exchanged with the external  periodic driving.\\
Repeating the construction of the weak coupling generator in the interaction picture one obtains
\begin{equation}\label{decomposition1}
\mathcal{L} = \sum_{\alpha}\sum_{\{\bar{\omega}_q \}_+}\mathcal{L}^{\alpha}_{\bar{\omega}_q},
\end{equation}
with a single term defined as
\begin{align}\label{eq_LindbladFast}
\mathcal{L}^{\alpha}_{\bar{\omega}_q}\rho	 = &\frac{1}{2}\Bigl\{G_{\alpha}(\bar{\omega}_q) \Bigl( [S_{\alpha}(\bar{\omega}_q), \rho S_{\alpha}(\bar{\omega}_q)^{\dagger}] + 
[S_{\alpha}(\bar{\omega}_q) \rho,  S_{\alpha}(\bar{\omega}_q)^{\dagger}]  \Bigr) \nonumber\\
& +G_{\alpha}(-\bar{\omega}_q)\Bigl( [S_{\alpha}(\bar{\omega}_q)^{\dagger}, \rho S_{\alpha}(\bar{\omega}_q)] + 
[S_{\alpha}(\bar{\omega}_q)^{\dagger} \rho,  S_{\alpha}(\bar{\omega}_q)]  \Bigr)\Bigr\}.
\end{align}
Using \eqref{SrelationsP} one can show that  $\mathcal{L}^{\alpha}_{\bar{\omega}_q}$ commutes with $-i[\bar{H},\cdot]$ and possesses a Gibbs-like stationary state
\begin{equation}\label{Gibbst}
\bar{\rho}^{\alpha}_{\bar{\omega}_q}= Z^{-1} e^{- (\bar{\omega}_q/\bar{\omega})\bar{H}/T_{\alpha}(\bar{\omega}_q)}, 
\end{equation}
with the local temperature $ T_{\alpha}(\bar{\omega}_ q)$ corresponding to the \emph{coupling channel} $(\alpha,\bar{\omega}_q)$. The ``renormalizing" term  ${\bar{\omega}_q}/{\bar{\omega}}$ in front of $\bar{H}$ in the Gibbs-like state takes into account the total energy exchange including $q\Omega$ corresponding to external driving device. The properties \eqref{SrelationsP} imply that the generator \eqref{eq_LindbladFast} transforms independently the diagonal and off-diagonal elements of $\rho$ computed in the eigenbasis of  $\bar{H}$.
\par
The MME the \emph{Schroedinger picture}  possesses the following structure
\begin{equation}\label{eq_MME_Schroedinger}
	\frac{d\rho_{sch}(t)}{dt} = -i \comm{H_S(t)}{\rho_{sch}(t)} + (\usup(t) \lgen \, \, \usup(t)^{\dagger}) \rho_{sch}(t),
\end{equation}
where  $\usup(t) \rho = U(t) \rho U(t)^{\dagger}$. A very useful factorization property for the solution of \eqref{eq_MME_Schroedinger} holds
\begin{equation}\label{eq_MME_Schroedingersol}
	\rho_{sch}(t) = \usup(t) e^{t\lgen} \rho_{sch}(0),
\end{equation}
which allows to discuss separately the decoherence/dissipation effects described by $\lgen$ and the unitary evolution  $\usup(t)$.
\subsection{The Laws of Thermodynamics}
The structure of MME's derived above and the  introduced notion of local temperatures allow to formulate the first and second law of thermodynamics in terms of energy, work, heat and entropy balance. The basic tool
is the following  inequality valid for any LGKS generator $\mathcal{L}$ with a stationary state $\bar{\rho}$ \cite{spohn2007irreversible} and arbitrary $\rho$
\begin{equation}
\mathrm{Tr}\bigl(\mathcal{L}\rho[\ln \rho - \ln \bar{\rho})]\bigr) \leq 0.
\label{spohn}
\end{equation}
This inequality is used to show positivity of entropy production under the assumption that the physical entropy of the system $S(t)$ is identified with the von Neumann entropy of its state
\begin{equation}
S(t)= -\mathrm{Tr}\bigl(\rho(t)\ln\rho(t)\bigr) .
\label{vNeumann}
\end{equation}
\subsubsection{Entropy balance and local heat currents}
In the case of fast driving there is no obvious definition of the temporal internal energy of the system because a fast exchange of energy quanta $q\Omega$ between system and the external source of driving makes a temporal partition of energy between both systems ambiguous. The situation is different for the entropy balance because the entropy change is due to irreversible processes which are slow under the weak system-environment coupling assumption. The weak coupling scheme yields the coarse-grained in time effective dynamics described by the MME
\eqref{eq_MME_Schroedinger} what suggest the following definition of the heat current $J_{\alpha}(\bar{\omega}_q)$ supplied to the system by the coupling channel $(\alpha, \bar{\omega}_q)$ and involving the averaged Hamiltonian (multiplied by  ${\bar{\omega}_q}/{\bar{\omega}}$)
\begin{equation}
J^{\alpha}_{\bar{\omega}_q}(t) =\frac{\bar{\omega}_q}{\bar{\omega}} \mathrm{Tr}\bigl(\bar{H}\mathcal{L}^{\alpha}_{\bar{\omega}_q}\rho(t)\bigr),
\label{heat current}
\end{equation} 
where $\rho(t)$ is the system density matrix in the interaction picture and according to \eqref{eq_MME_Schroedinger} given by
\begin{equation}\label{eq_MME_Schroedinger1}
	\rho(t) = e^{t\lgen} \rho(0).
\end{equation}
Those definition allow to formulate the Second Law which is again a consequence of \eqref{spohn} applied to each single coupling channel
\begin{equation}
\frac{d}{dt}S(t) - \sum_{\alpha}\sum_{\{\bar{\omega}_q \}_+}\frac{J^{\alpha}_{\bar{\omega}_q}(t)}{T_{\alpha}(\bar{\omega}_q)} = 
\sum_{\alpha}\sum_{\{\bar{\omega}_q \}_+}\sigma^{\alpha}_{\bar{\omega}_q}(t)\geq 0, 
\label{SIIlaw_fast}
\end{equation}
where $\sigma^{\alpha}_{\bar{\omega}_q}(t)\geq 0 $ is an entropy production caused by a single coupling channel $(\alpha, \bar{\omega}_q) $ given by
\begin{equation}
\sigma^{\alpha}_{\bar{\omega}_q}(t) = \mathrm{Tr}\bigl(\mathcal{L}^{\alpha}_{\bar{\omega}_q}\rho(t)[\ln \rho(t) - \ln \bar{\rho}^{\alpha}_{\bar{\omega}_q}]\bigr)\geq 0.
\label{Senprod_fast}
\end{equation} 
For the case of environment composed of several independent heat baths the equation \eqref{SIIlaw_fast} reduces to the standard form of the  Second Law for open systems with usual temperatures.

\subsubsection{Steady state regime }
Under natural ergodic conditions and due to \eqref{eq_MME_Schroedinger} any initial state tends to a limit cycle (or fixed point in particular cases), i.e.,
\begin{equation}
\rho(t) \to \bar{\rho}(t) = U(t) \bar{\rho}\, U(t)^{\dagger}= \bar{\rho}(t+\tau) , \quad\mathrm{where} \quad  \mathcal{L}\bar{\rho} = 0. 
\label{limit}
\end{equation}
Then the entropy $S(\bar{\rho}(t))$ and the heat currents given by
\begin{equation}
\bar{J}^{\alpha}_{\bar{\omega}_q} =\frac{\bar{\omega}_q}{\bar{\omega}} \mathrm{Tr}\bigl(\bar{H}\mathcal{L}^{\alpha}_{\bar{\omega}_q}\bar{\rho}\bigr),
\label{heat current_st}
\end{equation} 
become constants leading to the following form of the Second Law
\begin{equation}
\sum_{\alpha}\sum_{\{\bar{\omega}_q \}_+}\frac{\bar{J}^{\alpha}_{\bar{\omega}_q} }{T_{\alpha}(\bar{\omega}_q)} \leq 0.
\label{SIIlaw_fast1}
\end{equation}
The averaged internal energy of the system is constant in the limit cycle, and hence we can use the total energy conservation to write the First Law in the form
\begin{equation}
\bar{P}= -\sum_{\alpha}\sum_{\{\bar{\omega}_q \}_+}{\bar{J}^{\alpha}_{\bar{\omega}_q} },
\label{Ilaw_fast1}
\end{equation}
where $\bar{P}$ is the stationary power and if it is negative, it is supplied to the source of external driving.
\par
\textbf{Remark}\\
Because each $\mathcal{L}^{\alpha}_{\bar{\omega}_q}$ transforms diagonal (in $\bar{H}$ basis) elements of the density matrix into diagonal ones, the stationary state $\bar{\rho}$ is diagonal and hence the expressions for the stationary local heat currents and power involve only diagonal elements and the ``classical'' transition probabilities between them.
\subsection{ Carnot bound at steady state}
In the steady state regime the incoming and outgoing heat currents can be defined as follows
\begin{equation}
\bar{J}^{(+)} =  \sum_{\{\alpha,\{\omega_q\}_+ ; {\bar{J}}^{\alpha}_{\omega_q} > 0\}}{\bar{J}}^{\alpha}_{\omega_q},\quad \bar{J}^{(-)} =  \sum_{\{\alpha,\{\omega_q\}_+ ; {\bar{J}}^{\alpha}_{\omega_q} < 0\}}\bigl[-{\bar{J}}^{\alpha}_{\omega_q}\bigr].
\label{inout}
\end{equation}
We can introduce also effective ``hot/cold bath temperatures'' by averaging the inverse local temperatures with the weights proportional to incoming/outgoing heat currents
\begin{equation}
\frac{1}{T^{(+)}}= \sum_{\{\alpha,\{\omega_q\}_+ ; {\bar{J}}^{\alpha}_{\omega_q} > 0\}}\Bigl[\frac{{\bar{J}}^{\alpha}_{\omega_q}}{\bar{J}^{(+)}}\Bigr]\frac{1}{T_{\alpha}(\omega_q)}, \quad \frac{1}{T^{(-)}}= \sum_{\{\alpha,\{\omega_q\}_+ ; {\bar{J}}^{\alpha}_{\omega_q} < 0\}}\Bigl[\frac{{\bar{J}}^{\alpha}_{\omega_q}}{\bar{J}^{(-)}}\Bigr]\frac{1}{T_{\alpha}(\omega_q)}.
\label{hot_cold}
\end{equation}
Combining now \eqref{SIIlaw_fast1} - \eqref{hot_cold} and the standard notion of an efficiency of a heat engine  one obtains the generalized Carnot bound
\begin{equation}
\eta = \frac{-\bar{P}}{\bar{J}^{(+)}}\leq 1-\frac{T^{(-)}}{T^{(+)}},
\label{efficiency}
\end{equation}
which again coincides with the standard one in the case of environment composed of two heat baths.

\section{Conclusions}

We showed that quantum machines weakly coupled to  a single non-equilibrium stationary environment, and subject to fast periodic driving by work reservoirs, can be described by the thermodynamical principles and bounds which are very similar to the standard ones if only the proper definitions of the basic notions are used. In particular the notion of local temperature which depends not only on the state of environment but also on the form of system-environment coupling is crucial. 

The developed non-equilibrium theoretical framework may be used also to described the standard heat engine model, which operates under the interaction with two thermal baths. They can been effectively described as a single non-equilibrium bath. Therefore, we show that  standard heat engines are just  particular examples  of   non-equilibrium heat machines.

Starting from a bosonic thermal bath we showed how to obtain non-equilibrium baths by using different filters. Displacement or squezeeing operations may also be used. 
We found out that such non-equilibrium baths may be divided into two different types: (i)\textit{passive}, which equilibrate systems to different temperatures depending on their frequency but cannot drive heat engines, such baths can be obtained by  frequency independent transformations of equilibrium ones; (ii) \textit{non-passive}, which in addition allow  work extraction from a single bath and, in  the case of bosonic reservoirs, can be engineered by  frequency dependent transformations of equilibrium states.  They are also farther away from equilibrium than passive baths. 

A case where non-equilibrium bath is not stationary but, for example, is also perturbed by an external periodic driving is another interesting topic with possible applications.
A natural example is a spin-1/2 coupled to a spin-bath, both periodically perturbed by external magnetic field. It seems that the theory presented above can be extended to these cases as well.

\section*{Acknowledgements}

R.A. is supported by the Foundation for Polish Science TEAM project cofinanced by the EU European Regional Development Fund, and D. G-K by CONACYT.

%
%\bibliography{localtemp}

\end{document}